\documentstyle[12pt]{article}
\begin{document}
\title {Higher grade hybrid model of layered superconductors}
\author {by\\Ma{\l}gorzata Sztyren
\thanks{Department of Mathematics and Information Science,
Warsaw University of Technology, Pl. Politechniki 1, PL-00-661 Warsaw
E--mail: emes@mech.pw.edu.pl}}
\date \today

\input{epsf.tex}
\edef\undtranscode{\the\catcode`\_} \catcode`\_11
\newbox\box_tmp 
\newdimen\dim_tmp 
\def\jump_setbox{\aftergroup\after_setbox}
%
%
\def\resize
    #1
    #2
    #3
    #4
    {%
    \dim_r#2\relax \dim_x#3\relax \dim_t#4\relax
    \dim_tmp=\dim_r \divide\dim_tmp\dim_t
    \dim_y=\dim_x \multiply\dim_y\dim_tmp
    \multiply\dim_tmp\dim_t \advance\dim_r-\dim_tmp
    \dim_tmp=\dim_x
    \loop \advance\dim_r\dim_r \divide\dim_tmp 2
    \ifnum\dim_tmp>0
      \ifnum\dim_r<\dim_t\else
        \advance\dim_r-\dim_t \advance\dim_y\dim_tmp \fi
    \repeat
    #1\dim_y\relax
    }
\newdimen\dim_x    
\newdimen\dim_y    
\newdimen\dim_t    
\newdimen\dim_r    
\def\perc_scale#1#2{
  \def\after_setbox{%
    \hbox\bgroup
    \dim_tmp\wd\box_tmp \divide\dim_tmp100 \wd\box_tmp#1\dim_tmp
    \dim_tmp\ht\box_tmp \divide\dim_tmp100 \ht\box_tmp#2\dim_tmp
    \dim_tmp\dp\box_tmp \divide\dim_tmp100 \dp\box_tmp#2\dim_tmp
    \box\box_tmp 
\egroup}%
    \afterassignment\jump_setbox\setbox\box_tmp =
}%
{\catcode`\p12 \catcode`\t12 \gdef\PT_{pt}}
\def\hull_num{\expandafter\hull_num_}
\expandafter\def\expandafter\hull_num_\expandafter#\expandafter1\PT_{#1}
\def\find_scale#1#2{
  \def\after_setbox{%
    \resize\dim_tmp{100pt}{#1}{#2\box_tmp}%
    \xdef\lastscale{\hull_num\the\dim_tmp}\extra_complete}%
  \afterassignment\jump_setbox\setbox\box_tmp =
}
\def\scaleto#1#2#3#4{
  \def\extra_complete{\perc_scale{#3}{#4}\hbox{\box\box_tmp}}%
  \find_scale{#1}#2}
\let\xyscale\perc_scale
\def\zscale#1{\xyscale{#1}{#1}}
\def\yxscale#1#2{\xyscale{#2}{#1}}
\def\xscale#1{\xyscale{#1}{100}}
\def\yscale#1{\xyscale{100}{#1}}
\def\xyscaleto#1{\scaleto{#1}\wd\lastscale\lastscale}
\def\yxscaleto#1{\scaleto{#1}\ht\lastscale\lastscale}
\def\xscaleto#1{\scaleto{#1}\wd\lastscale{100}}
\def\yscaleto#1{\scaleto{#1}\ht{100}\lastscale}
\def\slant#1{
  \hbox\bgroup
  \def\after_setbox{%
    \box\box_tmp 
\egroup}%
  \afterassignment\jump_setbox\setbox\box_tmp =
}%
\def\rotate#1{
  \hbox\bgroup
  \def\after_setbox{%
    \setbox\box_tmp\hbox{\box\box_tmp}
    \wd\box_tmp 0pt \ht\box_tmp 0pt \dp\box_tmp 0pt
    \box\box_tmp
    \egroup}%
  \afterassignment\jump_setbox\setbox\box_tmp =
}%
\newdimen\box_tmp_dim_a
\newdimen\box_tmp_dim_b
\newdimen\box_tmp_dim_c
\def\plus_{+}
\def\minus_{-}
\def\revolvedir#1{
  \hbox\bgroup
   \def\param_{#1}%
   \ifx\param_\plus_ \else \ifx\param_\minus_
   \else
     \errhelp{I would rather suggest to stop immediately.}%
     \errmessage{Argument to \noexpand\revolvedir should be either + or -}%
   \fi\fi
  \def\after_setbox{%
    \box_tmp_dim_a\wd\box_tmp
    \setbox\box_tmp\hbox{%
     \ifx\param_\plus_\kern-\box_tmp_dim_a\fi
     \box\box_tmp
     \ifx\param_\plus_\kern\box_tmp_dim_a\fi}%
    \box_tmp_dim_a\ht\box_tmp \advance\box_tmp_dim_a\dp\box_tmp
    \box_tmp_dim_b\ht\box_tmp \box_tmp_dim_c\dp\box_tmp
    \dp\box_tmp0pt \ht\box_tmp\wd\box_tmp \wd\box_tmp\box_tmp_dim_a
    \kern \ifx\param_\plus_ \box_tmp_dim_c \else \box_tmp_dim_b \fi
    \box\box_tmp
    \kern -\ifx\param_\plus_ \box_tmp_dim_c \else \box_tmp_dim_b \fi
    \egroup}%
  \afterassignment\jump_setbox\setbox\box_tmp =
}%
\def\revolve{\revolvedir-}
\def\xflip{
  \hbox\bgroup
  \def\after_setbox{%
    \box_tmp_dim_a.5\wd\box_tmp
   \setbox\box_tmp
     \hbox{\kern-\box_tmp_dim_a \box\box_tmp \kern\box_tmp_dim_a}%
   \kern\box_tmp_dim_a
    \box\box_tmp
    \kern-\box_tmp_dim_a
    \egroup}%
  \afterassignment\jump_setbox\setbox\box_tmp =
}%
\def\yflip{
  \hbox\bgroup
  \def\after_setbox{%
    \box_tmp_dim_a\ht\box_tmp \box_tmp_dim_b\dp\box_tmp
    \box_tmp_dim_c\box_tmp_dim_a \advance\box_tmp_dim_c\box_tmp_dim_b
    \box_tmp_dim_c.5\box_tmp_dim_c
   \setbox\box_tmp\hbox{\vbox{%
     \kern\box_tmp_dim_c\box\box_tmp\kern-\box_tmp_dim_c}}%
   \advance\box_tmp_dim_c-\box_tmp_dim_b
   \setbox\box_tmp\hbox{%
     \lower\box_tmp_dim_c\box\box_tmp
}%
    \ht\box_tmp\box_tmp_dim_a \dp\box_tmp\box_tmp_dim_b
    \box\box_tmp
    \egroup}%
  \afterassignment\jump_setbox\setbox\box_tmp =
}%
\catcode`\_\undtranscode

\special{ps:}

\maketitle

\newcommand {\eqn}[1]{\begin{equation}#1\end{equation}}
\newcommand {\eqna}[1]{\begin{eqnarray}#1\end{eqnarray}}
\newcommand{\equln}[2]{\begin{equation} {#1} \label{#2} \end{equation}}
\newcommand {\la}{\longrightarrow}
{\bf Summary}
A hybrid discrete-continuous model of layered superconductors
with interlayer Josephson couplings of arbitrary range is
constructed. The conditions required by gauge invariance and
thermodynamical stability of the model are determined. Some
important special cases, in particular transition to the classic
Lawrence-Doniach model, are discussed. The conditions for
presence of alternating solutions and the posibilities to
describe such states of a superconductor by the continuum
or bi-continuum models are examined. The enhancement of
superconductivity caused by the presence of higher order
Josephson couplings is shown.



\section{Introduction}
   Most of the high-temperature superconductors like e.g. YBCO
or BSCCO have a layered structure. Such a strongly
anisotropic situation results in the fact that the material
properties and behaviour of the fields in direction (say \(z\) axis)
orthogonal to the layers is totally different from the behaviour
in directions paralel to layers. \cite{Blatter+ 94,Abrikosov
book,Rogula 99}.
The idea originally
proposed by Lawrence 
and Doniach (LD) \cite{Lawrence+Doniach 71} is to consider the
Ginzburg-Landau order parameter \(\psi\) as a function of two
continuous variables (say \(x\) and \(y\)) and one discrete variable
\(n\) - the index of the layer. The form of the free-energy
functional proposed in \cite{Lawrence+Doniach 71} and, after a
modification, presented e.g. in \cite{Ketterson+Song} takes into
account the Josephson coupling between the nearest neighbour layers.
The exact solutions for this case are given in \cite{KuplevakhskyB 2001}.
The higher grade hybrid model, proposed in this paper, admits also
couplings between
more distant neighbours - up to a (given, but arbirary) range \(K\).

   Let us look at the layered superconductor
as a one-dimensional chain of atomic planes
with Josephson's bonds between them. Such bonds
will be called J-links. The interplanar distance will be denoted
by \(s\). We assume
the following convention for indexing the planes and links.
If we locate the point \(z=0\) at an atomic plane,
then the \(z\)-coordinate of
any plane, equal \(ns\), may be represented by the integer \(n\),
while the \(z\)-coordinate of the center of any interplanar gap,
equal \(ls\),
by the half-integer \(l\). Choosing the point \(z=0\) at the center
of an interplanar gap - we index the planes by half-integers and the
gaps by integers.

\section{The hybrid model of grade K}
   Let us consider the free-energy functional \(\cal{F}\) for
a layered superconductor. We shall denote by \(\psi_n\) the order
parameter associated to the layer indexed by the number \(n\). Its
complex conjugate (c.c.) will be denoted by \(\bar{\psi}_n\).
The symbol \(m_{ab}\) and \(m_c\) will denote the in-plane and
tunneling effective mass of superconducting current carriers, respectively.
We start from the free energy functional of the following form 
\eqn{{\cal{F}}={\cal{F}}_0+{\cal{F}}_s+\frac{1}{8{\pi}}\int {\bf B}^2d^3x.}      
The term \({\cal F}_0\) describes the normal state, while \({\cal F}_s\)
the superconducting one. The supercoducting term is composed
of two parts: 
\equln{{\cal{F}}_s={\cal{F}}_p+{\cal{F}}_J,}{fs}
where the part
\eqn{{\cal{F}}_p=\sum_nF_n}
describes the contribution of atomic planes, while
\({\cal{F}}_J\) corresponds to interplanar Josephson's bonds.
For any plane indexed by \(n\) the free energy
\(F_n\) has the 2D Ginzburg-Landau form
(in general, the parameters can depend on \(n\))
\equln{F_{n}=s\int dxdy\{\frac{\hbar^2}{2m_{ab}}|({\bf D}\psi)_n|^2 
 +\alpha_0|\psi_n|^2+\frac{1}{2}\beta|\psi_n|^4\},}{efn}
where we have introduced the 2-dimensional continuous operator \({\bf D}\)
(covariant derivative)
\eqn{D_{\rho}=\partial _{\rho}-\frac{ie^*}{\hbar c}A_{\rho},
 \ \ \ \rho=x,y.}

The form (\ref{efn}) of the functional \({\cal{F}}_p\) already 
ensures its invariance with respect to the gauge transformation
\equln{{\bf A}\rightarrow {\bf A}'={\bf A}+\nabla\Lambda,\ \ 
 \psi_n\rightarrow \psi'_n=\psi_ne^{i\frac{e^*}{\hbar c}\Lambda},\ \ 
 \bar{\psi}_n\rightarrow \bar{\psi}'_n=
 \bar{\psi}_ne^{-i\frac{e^*}{\hbar c}\Lambda}}{gau}
(2-dimensional {\bf A} and \({\bf \nabla}\) for this case). The standard variational treatment
of the functional \({\cal F}_s\) with
respect to \(A_{\rho},\ \ \rho=x,y\), gives
the standard in-plane components of the supercoducting current
\equln{j_{\rho}=-\frac{ie^*\hbar}{2m_{ab}}(\bar{\psi}_n\partial_{\rho}\psi_n
     -\psi_n\partial_{\rho}\bar{\psi}_n)-\frac{e^{*2}}{m_{ab}c}A_{\rho}|\psi_n|^2
           ,\ \ \rho=x,y.}{incur}

   Let us now construct the term \({\cal{F}}_J\) in the free-energy.
In general it is a functional which can depend on all
\(\psi_n, \bar{\psi}_n\) and
on the vector potential \({\bf A}\). Consider
first the global gauge
transformation, i.e. \(\Lambda=const\) in (\ref{gau}).
The invariance condition for 
\({\cal{F}}_J\) implies that its density \(F_J\) fulfills the relation 
\eqn{\frac{\partial F_J}{\partial \psi_n}\psi_n-
   \frac{\partial F_J}{\partial \bar{\psi}_n}\bar{\psi}_n=0}
for each \(n\) separately. That means that the functional \({\cal{F}}_J\)
depends on the fields \(\psi\) only through the combinations
\(\bar{\psi}_n\psi_k\).

   Now we shall consider the local
gauge transformation with \(\Lambda\) depending only on the 
variable \(z\), but first let us introduce the following new variables
\eqn{\hat{\psi}_n=\psi_ne^{ip_{1n}}}
(and the appropriate complex conjugate), with
\eqn{p_{1n}=\frac{e^*}{\hbar c}\int_{ns}^{(n+1)s}A_zdz.}
The vanishing of the variation of \({\cal{F}}_J\) with respect to
\(\Lambda\) gives the condition
\eqn{\sum_n\frac{\delta F_J}{\partial \hat{\psi}_n}
 \frac{\partial\hat{\psi}_n}{\partial\Lambda}+c.c. +
 \frac{\partial F_J}{\partial A_z}
   \frac{\partial A_z}{\partial \Lambda}=0,}
which, together with the condition of global gauge invariance,
implies explicit independence of \({\cal{F}}_J\) of \(A_z\).\\
\indent
   The simplest gauge invariant expression for the energy of J-link
between \(n\)-th and (\(n+q\))-th planes will have the following form
\equln{\epsilon_{qn}=\frac{1}{2}\{\zeta_{qn}
 \bar{\psi}_n\psi_n+\eta_{qn}
 \bar{\psi}_{n+q}\psi_{n+q}-(\gamma_{qn}
 \bar{\psi}_n\psi_{n+q}e^{-ip_{qn}}+c.c.)\},}{eps1}
where the exponent \(p_{qn}\) is defined by the formula
\equln{p_{qn}=\frac{e^*}{\hbar c}\int_{ns}^{(n+q)s}A_zdz.}{pqn}
%
%
%
Let us note that for the model invariant with respect to the time
reversal we have \(\gamma_{qn}=\bar{\gamma}_{qn}\).
In general, the coupling parameters \(\zeta,\ \eta,\ \gamma\)
as well as in-plane parameters \(\alpha_0\) and \(\beta\) can be
different for different planes (which is the case for superconductors
composed of various atomic planes). However, in this paper we shall
consider the array of identical planes, hence the parameters will not
depend on index \(n\). That implies that \(\eta_q=\zeta_q\). Hence
instead of (\ref{eps1}) we shall use
\equln{\epsilon_{qn}=\frac{1}{2}\{\zeta_q
(|\psi_n|^2+|\psi_{n+q}|^2)
-\gamma_q
 (\bar{\psi}_n\psi_{n+q}e^{-ip_{qn}}+c.c.)\}.}{epsi}
Let \(P\) denote the (finite or infinite, but ordered) set of all
indices of planes, and let \(Q=\{1, 2,..., K\}\).
The planes connected with the \(n\)-plane by Josephson
coupling select the following subset of \(Q\)
\equln{P_n=\{q\epsilon Q:\ (n+q)\epsilon P\}.}{pn}
Thus, the gauge invariant functional \({\cal{F}}_J\)
for the hybrid model of grade K has the density
of the following form
\equln{F_J=\sum_{n\epsilon P}\sum _{q\epsilon P_n}\epsilon_{qn},}{fj}
with \(\epsilon_{qn}\) given by (\ref{epsi}).
The coupling parameters \(\zeta_q\) and \(\gamma_q\) vanish
for \(q>K\). Every J-link is represented in (\ref{fj}) by exactly
one term.

\section{Comparison with anisotropic GL model}

To compare our hybrid model (HM) with the continuum GL model,
we shall consider the infinite medium. In that case the summation
index may be shifted by the integer \(q\).
Moreover \(P_n=Q\).
This implies that
\equln{F_J=\frac{1}{2}\sum_n\sum_q\{2(\zeta_q-\gamma_q)|\psi_n|^2
 +\gamma_q|\psi_{n+q}e^{-ip_{qn}} -\psi_n|^2\}}{FJ}
For very weak field \(A_z\) and very small dependence of \(\psi_n\)
on \(n\) we have the correspondence rules which allows us
to pass from the hybride to the continuum models.
\eqn{
\left\{ \begin{array}{l}
\sum_n \rightarrow \frac{1}{s}\int dz,\\
\\
\frac{1}{qs}(\psi_{n+q}-\psi_n) \rightarrow \psi'_n(z),\\
\\
e^{-ip_{qn}} \rightarrow 1-i\frac{e^*}{\hbar c}qsA_z.\\
\end{array}
\right .}
Applying the rules to the functional (\ref{fs}) with (\ref{efn})
and (\ref{FJ}) one obtains
\begin{eqnarray}
{\cal F}_s \rightarrow \int d^3x\{\frac{\hbar^2}{2m_{ab}}
 |{\bf D}\psi|^2
 +\alpha_0|\psi|^2+\frac{1}{2}\beta|\psi|^4+\nonumber\\
 +\sum_q[(\zeta_q-\gamma_q)|\psi|^2+\frac{1}{2}q^2s^2\gamma_q|D_3\psi|^2]\}
.\end{eqnarray}

Hence, we have the following relation between \(m_c\) -- the effective
mass in \(z-\)direction (in anisotropic GL model) and the coupling
parameters \(\gamma_q\):
\eqn{\frac{1}{m_c}=\frac{s^2}{\hbar^2}\sum_qq^2\gamma_q.}
The presence of J-links modifies also the parameter \(\alpha_0\)
to the form:
\equln{\alpha=\alpha_0+\sum_q(\zeta_q-\gamma_q).}{alf}
\section{Field equations}

By computing the variation of the functional \({\cal F}\) with
respect to \(A_z\) one obtains the Maxwell equation for the
\(z-\)components
of current density and \({\rm curl}\, H\)
\eqn{\frac{1}{c}J(z)=\frac{1}{4{\pi}}({\rm curl}\, H)_z,}
where 
\equln{J(z)=i\frac{se^*}{2\hbar c}
\sum_{n\epsilon P}\sum_{q\epsilon P_{n}}
 \{\gamma_q\bar{\psi}_n\psi_{n+q}e^{-ip_{qn}}-c.c.\}\chi_{qn}(z),}{cn}
\(P_n\) is given by (\ref{pn}), and the quantity \(p_{qn}\) by (\ref{pqn}).
The symbol \(\chi_{qn}(z)\) denotes
the characteristic function of the interval \([ns,(n+q)s]\). Let us note
that for any layer \(ns<z<(n+1)s\) the exppression \(J(z)\) does not
depend on the value z; only the ends of the interval are important. To
better see the structure, let us first extend the set \(Q\)
on the negative values
\equln{\bar{Q}=\{-K,...,-2,-1,1,2,...,K\},}
{qbar}
and introduce the symbols
\equln{\sigma_{qn}=\left\{ \begin{array}{l}
1\ \ if\ \ \ (n+q)\ \epsilon P\ ,
\\
0\ \ otherwise\ .
\end{array}
\right .} {pn}
Then the expression for Josephson current \(J_l\) describing tunneling
across the interplanar gap indexed by \(l\) (half integer if \(P\)
contains integers) will have the form
\equln{J_l=i\frac{se^*}{2\hbar c}\sum_{q\epsilon \bar{Q}}
\sum_{n\epsilon P_{lq}}
 \{\gamma_q\sigma_{qn}\bar{\psi}_n\psi_{n+q}e^{-ip_{qn}}-c.c.\},}{cn}
where
\eqn{P_{lq}=\{n\epsilon\ P: n<l<n+q\}.}
By computing the variation of the functional
 \({\cal F}_s\) with
respect to \(\bar{\psi}_n\), one obtains the equations

\equln{
\begin{array}{l}
  -\frac{\hbar^2}{2m_{ab}}{\bf D}^2\psi_n+\tilde{\alpha}\psi_n+
     \beta|\psi_n|^2\psi_n+\nonumber\\\\
    -\frac{1}{2}\sum_{q\epsilon \bar{Q}}
    \gamma_q
    (\sigma_{qn}\psi_{n+q}e^{-ip_{qn}}+
    \sigma_{-qn}\psi_{n-q}e^{ip_{q,n-q}})=0,
\end{array}
}{eqpsi}
where instead of \(\alpha_0\) we have introduced
\equln{\tilde{\alpha}=\alpha_0+\sum_q\sigma_{qn}\zeta_q}{tilal}
depending of \(n\), for finite \(P\).

\section{The ground states}

    Let us now consider the plane-uniform states of HM in
the absence of magnetic field. The order parameter is then
independent of the in-plane variables and the net supercurrents
vanish. In detailed calculations we shall focus our attention
on the grade \(K=2\), which seems to be sufficiently illustrative
to grasp the idea on what is going on. The generalization to
grades \(K>2\), although more complicated algebraically,
is straightforward. 
In the region far from the boundary all the coefficients \(\sigma_{qn}=1\).
    For \(K=2\) the condition of vanishing Josephson current
is equivalent to
\eqn{\gamma_1(\bar\psi_n\psi_{n+1}-c.c.)
 +\gamma_2(\bar\psi_n\psi_{n+2}
 +\bar\psi_{n-1}\psi_{n+1}-c.c.)=0.}
and the equations (\ref{eqpsi}) take the form
\equln{\tilde{\alpha}\psi_n +\beta|\psi_n|^2\psi_n-\frac{1}{2}[\gamma_1
(\psi_{n+1}+\psi_{n-1})+\gamma_2(\psi_{n+2}+\psi_{n-2})]=0.}{far}
We shall look for  solutions with constant amplitude and
diference of phase between nieghbouring atomic planes, so we use the
ansatz
\eqn{\psi_n=Ce^{in\theta}.}
The result is the equation
\equln{\tilde{\alpha}+\beta C^2-\gamma_1 \cos\theta-\gamma_2\cos 2\theta=0,}
 {upsi}
with the condition
\equln{\gamma_1 \sin\theta+2\gamma_2 \sin 2\theta=0.}{teta}
Solving (\ref{teta}) with respect to \(\theta\), we obtain 3 variants:
(a1) \(\theta=0\), (a2) \(\theta={\pi}\), and
(a3) \(\cos\theta=-(\gamma_1//4\gamma_2)\). The solution \(C\) to (\ref{upsi})
has the form \(C^2=-\alpha^*//\beta\), where \(\alpha^*\) depends
on the variant.
The variant (a1) implies the uniform solution to (\ref{far}):
\equln{\psi_n=C,}{const}
with
\equln{\alpha^*=\alpha_0 + \zeta_1 + \zeta_2
-\gamma_1 - \gamma_2,}{al1}
what is the case of equation (\ref{alf}) for grade K=2.
In the variant (a2) the solution to (\ref{far}) has the alternating form
\equln{\psi_n=\pm C,}{alt}
with  
\equln{\alpha^*=\alpha_0 + \zeta_1 + \zeta_2
+\gamma_1 - \gamma_2.}{al2}
Finally, in the variant (a3), the solution exists if the parameters
\(\gamma_1\) and \(\gamma_2\) fulfill the relation
\equln{|\gamma_1|\,{{\leq}}4|\,\gamma_2|.}{gamy}
Then the parameter \(\alpha^*\) is connected with the coupling
constants by the formula:
\equln{\alpha^*=\alpha_0+\zeta_1+\zeta_2+\gamma_2(1+\frac{\gamma_1^2}
  {8\gamma_2^2}).}{al3}
There are two independent solutions
\equln{\psi_n=Ce^{\pm\theta},\ \ 
  \theta={\rm arccos}(-\frac{\gamma_1}{\gamma_2}).}{modu}
They will be referred to as the phase modulated states.
The solutions degenerate at the extremities
\(|\gamma_1|\,=4|\,\gamma_2|\).

    So far we confined our discussion to the existence of
solutions which could serve as candidates for the ground state.
The question of stability of the solutions
will be addressed in the next section.

    Let us note that the condition \(\sin\theta=0\) admits
the solutions (\ref{const}) and (\ref{alt}) for
hybrid model of any grade K. Contrary to that, the analogues of
the conditions (a3) and (\ref{gamy}) can deliver, depending on the grade
and the coupling parameters, any number from \(0\) to \(2(K-1)\)
modulated solutions.

\section{Stability}
To examine the stability of the solutions presented in the
previous section, we shall analyze the Hessian matrix of the 
free energy \({\cal F}_s\) describing small deviations from the ground state.
The problem reduces to examining the function
\equln{{\cal E}(C,\theta)=\tilde{\alpha}C^2+\frac{1}{2}\beta C^4
 -C^2(\gamma_1 \cos \theta+\gamma_2 \cos 2\theta).}{stab}
The solutions found in the previous section are stationary points
of \({\cal E}\).
If \(\gamma_2=0\), then the stability of the solutions depends on
the sign of \(\gamma_1\). If \(\gamma_2\, {\neq}\, 0\) then 
the sign of the respective eigenvalue
depends on the values of the parameters \(\gamma_1\)
and \(\gamma_2\). For \(\theta\) different from \(0\) and \({\pi}\)
the dependence has form according to the function
\equln{f(\gamma_1,\gamma_2)=\gamma_2(\gamma_1+4\gamma_2)(\gamma_1-4\gamma_2)
.}{deter}
The straight lines \(\gamma_1+4\gamma_2=0\) and \(\gamma_1-4\gamma_2=0\)
divide the plane \(\gamma_1,\gamma_2\) into 4 regions (see Figure 1).\\
\indent As explained above, both the uniform and the
alternating solutions always exist. However, in the region 
\equln{(A):\ \gamma_1 >0,\ -\gamma_1<4\gamma_2<\gamma_1,}{rA}
only the uniform solution (\ref{const}) is stable.\\\\
\vskip 0pt
\noindent \hfil\hbox{\epsffile{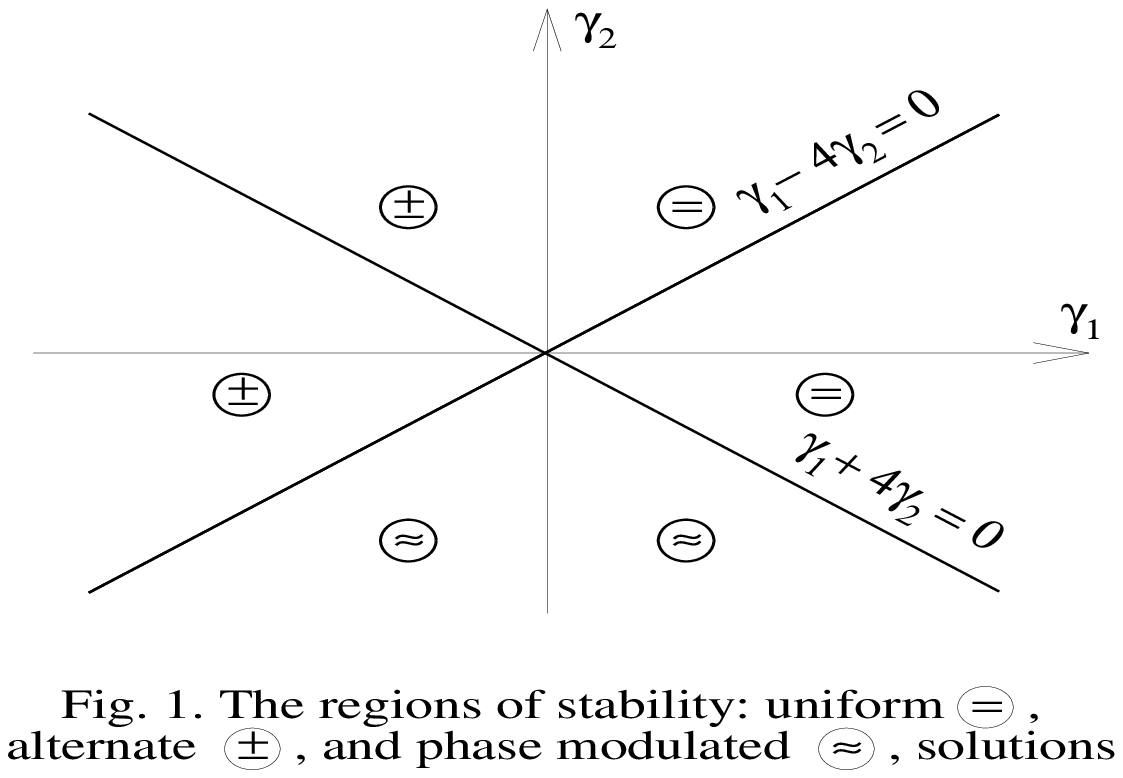}}\hfil\\
\vskip -0.5\baselineskip
\noindent
On the other hand,
in the region
\equln{(B):\ \gamma_2 >0,\ -4\gamma_2<\gamma_1<4\gamma_2,}{rB}
both the solutions
(\ref{const}) and (\ref{alt}) are stable (the uniform one in the right
and the alternating one in the left part of the region), while in the region
\equln{(C):\ \gamma_1 <0,\ -\gamma_1<4\gamma_2<\gamma_1,}{rC}
only the alternating solution (\ref{alt}) is stable. 
    The region
\equln{(D):\ \gamma_2 <0,\ 4\gamma_2<\gamma_1<-4\gamma_2,}{rD}
excludes the stability of both the uniform and the alternating
solutions
but, in contrast to that, ensures the existence and stability of
the modulated solutions (\ref{modu}).
In the region (\ref{rB}) the modulated solutions exist
but are unstable. 

   Let us note that in the regions (\ref{rB}) and (\ref{rC}) of
stability of the alternating solution (\ref{alt}) one can apply
the construction of bi-continuum solution presented in
\cite{Sztyren 2002}.

\section{Enhancement of the superconductivity}

The asociation of the formulae (\ref{al1}), (\ref{al2}) and (\ref{al3})
with the regions of stability of the ground state shows that,
for suitable relations between the coupling constants
\(\gamma_1\) and \(\gamma_2\), one can make the parameter
\(\alpha^*\) more negative than \(\alpha_0\). In consequence,
the 3D superconductivity can turn out enhanced with respect to
the 2D superconductivity in the atomic planes. Such a possibility
has been indicated in literature \cite{Anderson 92}.
  In further discusion we shall count the enhancement with respect
to the origin located at \(\tilde{\alpha}\) given by (\ref{tilal}). 
It is convenient to introduce the polar coordinates \(\gamma\)
and \(\varphi\) in the plane \(\gamma_1,\ \gamma_2\):
\eqn{\gamma_1=\gamma\cos\varphi,\ \gamma_2=\gamma\sin\varphi,}
and the notation
\eqn{\varphi_0=\arctan(\frac{1}{4}).}
\vskip 0pt
\noindent \hfil\hbox{\epsffile{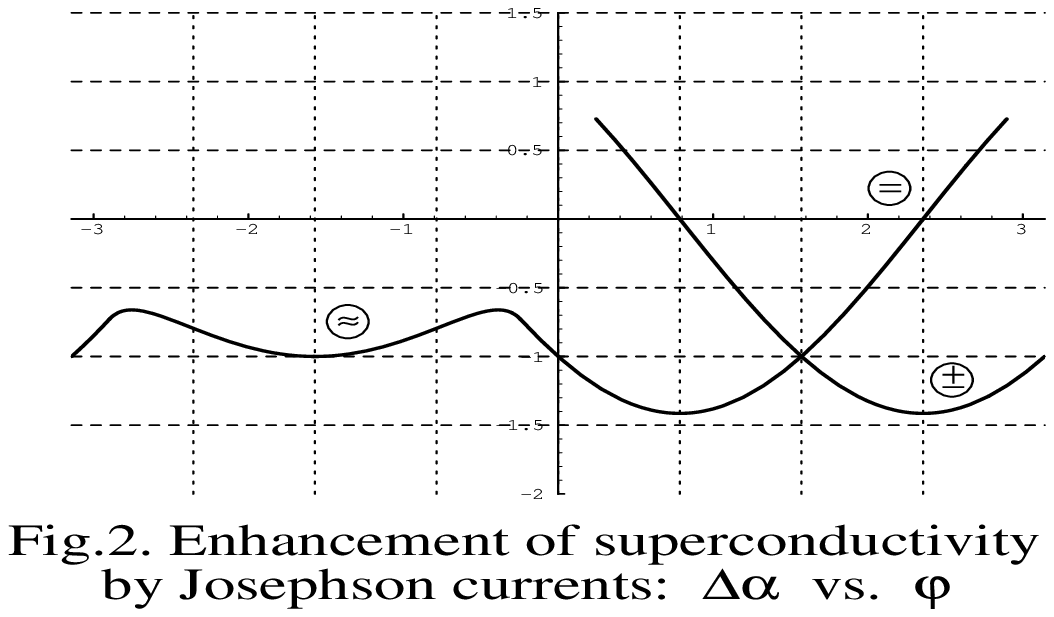}}\hfil\\
\vskip -0.5\baselineskip
The quantity 
\(\Delta\alpha=\alpha^*-\tilde{\alpha}\) as a function of the
coupling angle \(\varphi\) is plotted in Figure 2
(the numerical values are computed for
\(\gamma=1\)).\\
  In the uniform state (\ref{al1}) we have
\eqn{\Delta\alpha=\sqrt{2}\gamma\sin(\varphi-\frac{3}{4}{\pi}), 
 \ -\varphi_0\,{{\leq}}\varphi\,{{\leq}}{\pi}-\varphi_0.}
The minimum \(\Delta\alpha\) (hence the maximum enhancement)
is reached at \(\varphi={\pi}//4\) and equals \(-\sqrt{2}\gamma\).

   In the alternating state (\ref{al2}), in turn, one obtains
\eqn{\Delta\alpha=\sqrt{2}\gamma\sin(\varphi+\frac{3}{4}{\pi}),
 \ +\varphi_0\,{{\leq}}\varphi\,{{\leq}}\,{\pi}+\varphi_0}
with the minimum value \(-\sqrt{2}\gamma\) reached at 
\(\varphi=\frac{3}{4}{\pi}\).

   The enhancement for the phase modulated state (\ref{al3})
is represented by
\eqn{\Delta\alpha=\gamma\sin\varphi(1+\frac{1}{8}{\rm ctg}^2\varphi),
 \ -{\pi}+\varphi_0\,{{\leq}}\,\varphi \,{{\leq}}\,-\varphi_0.}
In this case the minimum equals \(-\gamma\) and is reached at \(\varphi=-{\pi}//2\).
Hence, the maximum enhancement in this case is smaller than
in the uniform and alternating states.

\section{Special cases}

   The hybrid model of grade 0 (HM with K=0) reduces formally to uncoupled
(2 dimensional) Ginzburg-Landau
model for each atomic plane. The HM with K=1 has in general two
coupling parameters \(\zeta\) and \(\gamma\). If they are equal to
one another, one obtains the Lawrence-Doniach model
with parameter \(\alpha=\alpha_0\), and the effective mass
in z--direction given by the formula
\eqn{\frac{1}{m_c}=\frac{s^2}{\hbar^2}\gamma.}
   Due to \(\gamma>0\) the LD model has only one stable ground state
solution, namely the uniform one.
The interplanar coupling
gives neither enhancement nor suppression of the
critical temperature. Although the Josephson current coupling
places the model on the enhancement side, the efect is
precisely annuled by \(\zeta=\gamma\).

   Let us note that, in the enhancement mechanism discussed above,
the 2D superconductivity of the planes is not a prerequsite for
the 3D superconductivity of the array of the planes. In fact, one can
obtain the negative \(\alpha^*\) starting from
positive \(\alpha_0\). This is in concordance with ideas expressed
in Anderson's discussion of his Dogma V in \cite{Anderson 97 book}.

\section{Acknowledgements}
This work was supported by the Science Research
Committee (Poland) under grant No. 5 TO7A 040 22.\\

\thebibliography{12}
\bibliography{}

\bibitem{Abrikosov book}A.\,A. Abrikosov,
Fundamentals of the Theory of Metals, North--Holland 1988  

\bibitem{Anderson 92}P.\,W. Anderson,
{\em Intralayer tunneling mechanism for high--T\(_c\)   superconductivity: comparison with c-axis infrared experiments}   included in: P.W.Anderson, The theory of superconductivity in   the high--$T_c$ cuprates, Princeton University Press 1997  

\bibitem{Anderson 97 book}P.\,W. Anderson,
{\em The theory of superconductivity in the high--$T_c$    cuprates,} Princeton University Press 1997  

\bibitem{Blatter+ 94}G. Blatter, M.\,V. Feigel'man, V.\,B. Geshkenbein and A.\,I. Larkin and V.M. Vinokur,
{\em Vortices in high--temperature superconductors,}    Reviews of Modern Physics {\bf 66}, 1125(1994)  

\bibitem{Ketterson+Song}J.\,B. Ketterson and S.N. Song,
Superconductivity,       Cambridge Univ. Press 1999  

\bibitem{KuplevakhskyB 2001}S.\,V. Kuplevakhsky,
{\em "Vortices and the Meissner effect in stacked junctions     and layered superconductors: Exact analytical result,}    Phys. Rev. B {\bf 63}, 054508 (2001)  

\bibitem{Lawrence+Doniach 71}W.\,E. Lawrence and S. Doniach,
{\em Theory of layer structure superconductors,} in Proceedings of the Twelfth International Conference on Low Temperature Physics, Kyoto, ed. E.Kanda, Academic Press of Japan 1971, p.361  

\bibitem{Rogula 99}D. Rogula,
{\em Dynamics of magnetic flux lines and critical fields in high T\(_c\) superconductors}, Journal of Technical Physics {\bf 40} (1999), p.383  

\bibitem{Sztyren 2002}M. Sztyren,
{\em Bi-continuum modelling of layered structures and crystalline interfaces}, Journal of Technical Physics {\bf 43} (2002), p.265
\end{document}